%% file: paper.tex
\documentclass[pageno, inline]{jpaper}

\usepackage[normalem]{ulem}
\usepackage{amsmath}
\usepackage{amssymb}
\usepackage[draft=true]{minted}

\begin{document}

\title{Drndalo: Lightweight Control Flow Obfuscation Through Minimal
Processor/Compiler Co-Design}
\author{Novak Boskov, Mihailo Isakov, Michel A. Kinsy \\
Adaptive and Secure Computing Systems (ASCS) Laboratory\\
Department of Electrical and Computer Engineering, Boston University\\
\{boskov, mihailo, mkinsy\}@bu.edu \vspace{-0.1in}}

\date{}
\maketitle

\thispagestyle{empty}

\begin{abstract}
  Binary analysis is traditionally used in the realm of malware
  detection. However, the same technique may be employed by an
  attacker to analyze the original binaries in order to reverse
  engineer them and extract exploitable weaknesses. When a binary is
  distributed to end users, it becomes a common remotely exploitable
  attack point. Code obfuscation is used to hinder reverse engineering
  of executable programs. 
 In this paper, we focus on securing binary distribution,
  where attackers gain access to binaries distributed to end devices,
  in order to reverse engineer them and find potential vulnerabilities. 
  Attackers do not however have means to monitor the execution of said devices. 
  In particular, we focus on the control flow obfuscation --- a technique 
  that prevents an attacker from restoring
  the correct reachability conditions for the basic blocks of a
  program. By doing so, we thwart attackers in their effort to infer
  the inputs that cause the program to enter a vulnerable state (e.g.,
  buffer overrun). We propose a compiler extension for obfuscation and
  a minimal hardware modification for dynamic deobfuscation that takes
  advantage of a secret key stored in hardware. We evaluate our
  experiments on the LLVM compiler toolchain and the BRISC-V open
  source processor. On PARSEC benchmarks, our deobfuscation technique
  incurs only a 5\% runtime overhead. We evaluate the security of
  Drndalo by training classifiers on pairs of obfuscated and
  unobfuscated binaries. Our results shine light on the difficulty of
  producing obfuscated binaries of arbitrary programs in such a way
  that they are statistically indistinguishable from plain binaries.
\end{abstract}

\input{intro}

\input{distribution}

\input{obfuscate}

\input{deobfuscate}

\input{deobfuscation_results}
\input{evaluate}
\input{security_evaluation}
\input{related_work}
\input{future_work}
\input{concl}

\bibliographystyle{plain}
\bibliography{paper}

\end{document}

%% file: intro.tex
\section{Introduction}

There is a multitude of software vulnerabilities that allow an
attacker to gain unauthorized access to the critical parts of a
multi-user system. For example, in modern server environments, an
attacker logged in as a regular user is interested in running a
malicious program as the privileged user. One of the techniques to
achieve their goal is privilege escalation through exploiting buffer
overflows. To mount a buffer overflow, the attacker looks for an
appropriate location in the targeted binary. If the plain version of
the binary that runs in privileged mode is available to the attacker,
the job of hunting for the mount points becomes easier. For example,
the attacker may undertake the reverse engineering of the original
software or can only explore the input space to find the malicious
inputs. To prevent reverse engineering, software vendors resort to
software obfuscation --- a technique generally applicable to a wide
range of problems as we discuss in Section~\ref{sec:rel_work}.

One of the most effective obfuscation targets is the program's control
flow (CF).
CF obfuscation is a technique in which a dedicated program called the
obfuscator performs semantic-preserving transformations on the
original program in order to hide the original CF.\
This kind of obfuscation traditionally heavily relies on opaque
predicates. A predicate is opaque if its resolution is hard or
ambiguous for the attacker. The technique of opaque predicates is used
in obfuscation tools such as Obfuscator-LLVM~\cite{Junod_2015}. The
construction of an opaque predicate is done by tailoring a
computationally intensive challenge for the underlying concolic
execution engine such as present in Mayhem~\cite{Cha_2012},
Angr~\cite{Shoshitaishvili_2016} or
Triton~\cite{SSTIC2015-Saudel-Salwan}. Some of the challenges proposed
by Xu et. al.~\cite{Xu_2018} are: symbolic memory, floating-point
algebra, covert symbolic propagation and parallel programming. This
approach provides the mechanism for constructing multiple different
concrete challenges from the same basic templates. Despite a certain
amount of generality, an attacker can expose the constructed
challenges by observing similar patterns in the critical sections of
the binary. Upon successful detection of such patterns, an attacker
can unfold them to restore the original semantics. This is possible
because the majority of the challenges that are constructed for this
purpose have completely deterministic behavior. That is, for the
particular input they always produce the same output. But the power of
a versatile attacker goes far beyond that. New advances in concolic
execution engines allow them to solve more of computationally
intensive tasks generated by opaque predicates. The state-of-the-art
techniques for binary analysis of real world software are becoming
practical and
mature~\cite{Avgerinos_2014,Shoshitaishvili_2016,Cha_2012,Schulte_2018,Eagle:2011:IPB:2049962,ghidra,wang2015,Stephens_2016,SSTIC2015-Saudel-Salwan}. For
example, Angr~\cite{Shoshitaishvili_2016} is powerful enough to
automatically translate low level disassembler information to an
abstract level in which it can do unified analysis for a variety of
different platforms. It can also perform symbolic analysis of the
program's control flow to infer reachability conditions for the basic
blocks of interest. Furthermore, it can simulate the processor
execution and even the interaction with the operating system. When
symbolic analysis falls short, Angr supports different fuzzing
techniques~\cite{Stephens_2016} to help symbolic analysis. We discuss
the function of Angr in more details in Section~\ref{sec:angr}.

The advances in flexibility of hardware
design promoted by the initiatives such as RISC-V and OpenRISC create
fertile soil for hardware-software co-designs. However, RISC-V
  binaries pose easier targets for reverse engineering than, e.g., x86
  binaries. As opposed to x86 binaries, RISC-V binaries have
  fixed-size instruction lengths and clearly separate code from data,
  which is one of the challenges for reverse engineering of x86
  binaries. Thus, we rethink the CF obfuscation in the context of
hardware-software co-design and propose Drndalo --- a lightweight
hardware-assisted control flow obfuscation technique. Our approach
does not rely on the lack of
  obfuscation-specific features of the binary analysis frameworks nor
their theoretical limitations, but on a safely distributed
hardware-software shared secret.

Whole-executable encryption may pose a plausible alternative to CF
obfuscation. The device running the encrypted code can either decrypt
the executable in bulk, or just-in-time~\cite{10.1007/978-3-540-79104-1_7}.
While this secures the binary against distribution-time attacks, in order
to maintain similar performance compared to a device running unencrypted code,
the device may need to have hardware decryption modules and enough memory
to store the decrypted code. For small embedded devices, this may be
prohibitive due to hardware area (potentially increasing the device cost),
and power (both from the cryptographic accelerator and additional memory).
We therefore propose a lightweight alternative to binary encryption,
both in terms of area and power.

The main contributions of this paper are:
\begin{itemize}
\item We introduce a new hardware-assisted CF obfuscation technique
  utilizing a minimal extension to BRISC-V processor core, LLVM
  compiler and a secret key stored in hardware,
\item We evaluate the security of our method against automated attacks
  based on different classifiers and analyze their success rates,
\item We highlight and discuss the family of scenarios in which our
  obfuscation technique exhibits imperfections and propose possible
  enhancements.
\end{itemize}
The rest of the paper is organized as follows. In
Section~\ref{sec:attack_model} we model our attack scenario and
explain the means that an attacker has at its disposal. In
Section~\ref{sec:obf} we describe the details of the obfuscation
process that happens in the compiler. In Section~\ref{lab:deobf} we
explain the deobfuscation phase that takes place in hardware and
describe the needed extensions to underlying processor core. In
Section~\ref{sec:angr} we describe the internals of
Angr~\cite{Shoshitaishvili_2016} that empower the attacker from our
attack model. In Section~\ref{sec:eval} we evaluate the overhead of
our Drndalo method through a comparison of different deobfuscation
phase designs. In Section~\ref{sec:sec_eval} we evaluate the security
of our method against the classifiers that have access to obfuscated
and plain binaries. In Section~\ref{sec:rel_work} we give an overview
of the literature that tackles the problem of program
obfuscation. Finally, in Section~\ref{ref:future} we discuss the
possible extensions of our work to maximize its security against the
classification based attacks and conclude our paper in
Section~\ref{sec:concl}.

%% file: distribution.tex
\section{Attack Model}\label{sec:attack_model}
In our scenario, an organization developing software needs to deploy
the software to client machines, but wants to prevent attackers from
gaining access to the program's original behavior. The organization
does that by obfuscating reachability conditions for the basic blocks
in the compiled binaries. The organization performs the obfuscation
and sends the obfuscated version of the binary to clients. The
organization can (1) obfuscate binaries in a client-agnostic fashion,
or (2) can separately obfuscate the binaries for each individual
client. In either case, the binary is deobfuscated using some secret
value, e.g., a key or a physical unclonable function
(PUF)~\cite{4261134} challenge-response pair. Both obfuscation and
deobfuscation procedures are public and described in
Section~\ref{sec:obf} and Section~\ref{lab:deobf}, respectively. The
obfuscation hinges on the security of the secret value.

The attackers are able to steal any number of (possibly differently)
obfuscated binaries once they have left the organization, but are
unable to steal the deobfuscation keys. If the deobfuscation hinges on
a specific piece of hardware (as in the case of PUFs), the attackers
do not have access to the actual chip that can deobfuscate the
code. The attackers steal the obfuscated binaries by either
intercepting network traffic to the client or by stealing the binaries
from the drives of infected clients. The attackers can also perform
obfuscation of their own programs with their own keys arbitrarily many
times. In our scenario, the attackers cannot however monitor the
execution of the original binary on the target machines, i.e., the
binary theft and the binary execution occur at different times.

In our obfuscation and deobfuscation procedures that we discus in
Sections~\ref{sec:obf} and~\ref{lab:deobf} we use a cryptographic hash
function. In fact, the security of the hash function is orthogonal to
the Drndalo technique that we propose. In our experimental settings we
use a parametrizable Linear Feedback Shift Register (LFSR) due to its
minimal hardware cost. However, a defender with higher security
requirements may want to use a cryptographically secure hash function
(e.g., SHA) with higher hardware and performance costs.

%% file: obfuscate.tex
\section{Obfuscation Procedure}\label{sec:obf}
The essence of our approach relies on potentially inverting all the
conditional branches in the original program.
For each branch, the obfuscator decides whether to invert that branch
by evaluating a function that takes two inputs:
\begin{enumerate*}[label=(\arabic*)]
\item a unique identifier of the branch (e.g., the address of the
  instruction), and
\item a program key,
\end{enumerate*}
 and produces a 1-bit output.
If the function returns a 1, the branch condition is inverted,
otherwise it is not. In our implementation, we use a cryptographic
hash function with a binary output.
Hence, the only secret information is the deobfuscation key.

Table~\ref{tab:transitions} summarizes the translations that are used
for obfuscation and deobfuscation if the hash function answers with
the value 1 for the conditional branch under
consideration. 

\begin{figure}[h]
 \vspace{-0.2in}
  \centering
   \includegraphics[width=\columnwidth]{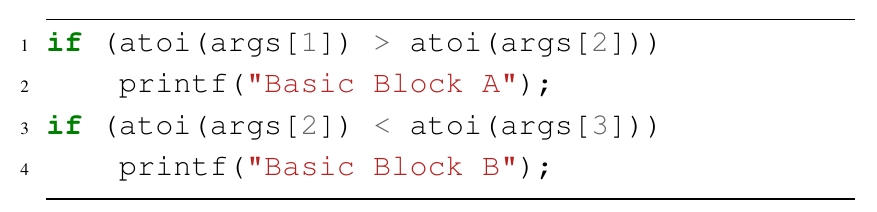}
  \caption{Segment of the example C program before obfuscation.}
  \label{lst:ccode}
\end{figure}

\begin{figure}[h]
\vspace{-0.2in}
  \centering
  \begin{minipage}[t]{.5\columnwidth}
    \definecolor{bg2}{rgb}{0.90,0.90,0.90}
   \includegraphics[width=\columnwidth]{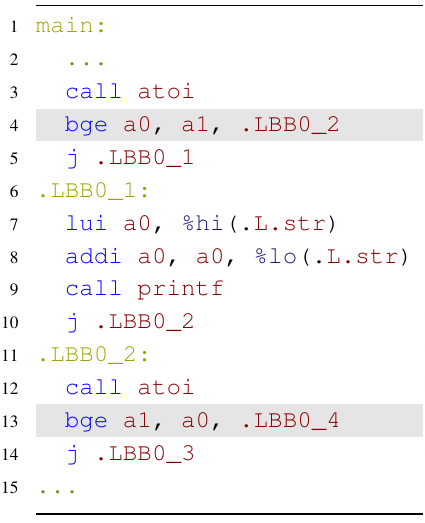}
    \caption{Plain code}
    \label{lst:plain}
  \end{minipage}%
  \begin{minipage}[t]{.5\columnwidth}
    \definecolor{bg2}{rgb}{0.95,0.95,0.95}
    \includegraphics[width=\columnwidth]{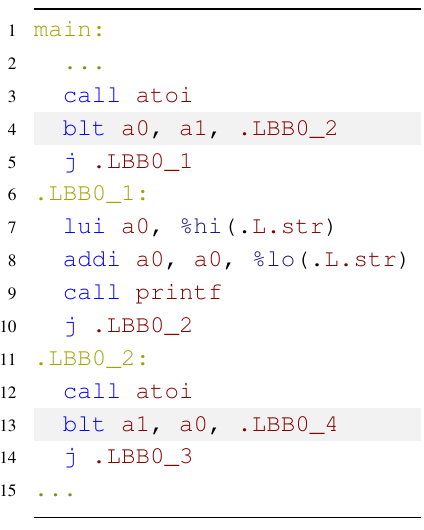}
    \caption{Obfuscated code}
    \label{lst:obf}
  \end{minipage}
  \caption{Example program assembly code before and after obfuscation.}
\end{figure}

Figure~\ref{lst:ccode} shows a segment of the original
code of an example C program while Figure~\ref{lst:plain} shows its
assembly.  Assuming that the hash function returns the value 1 for
both conditional branches marked in gray, the obfuscated RISC-V assembly is
shown in Figure~\ref{lst:obf}.

\input{latex_figures/transition_table}

The obfuscated code from Figure~\ref{lst:obf}, observed in isolation,
does not reveal that it is obfuscated even if our method is known to
the attacker. It represents a valid program in its own right and it
also does not contain obvious traces of obfuscation. That is, the
obfuscation is stealthy. In Section~\ref{sec:sec_eval} we use a
classifier to quantitatively evaluate the stealth of our Drndalo
obfuscation.

In contrast to our Drndalo method, some of the existing methods for
control flow obfuscation~\cite{Xu_2018} rely on the code snippets that
increase computational complexity of the original program. However,
those snippets, if known to the attacker, can be spotted in the
obfuscated program.

The main property that differentiates obfuscated program from the
plain program is the set of reachability conditions for the basic
blocks behind conditional branches. The topology of the obfuscated CF
graph and its original in the plain program are identical. When an
attacker executes the obfuscated program by providing it with the
inputs, the obfuscated program has significantly different behavior
from its original. That is, the traces of execution affected by the
reverted branches in the obfuscated program are different from the
traces in the plain program. The mutated runtime behavior will give
the attacker a wrong set of crashing inputs (see
Section~\ref{sec:angr}). Each conditional branch is either inverted or
not with the same probability (provided that the hash function is
secure). Given that
  the attacker does not apply semantic and symbolic analysis, the
  action of restoring the original program from the obfuscated program
  comes down to guessing the original branches from the obfuscated
  branches. If the attacker applies brute force only, its probability
  of success in restoring the original program from the obfuscated
  program is $P = 0.5^{n}$, where n is the number of conditional
  branches.

\begin{figure*}[!h]
  \centering
  \includegraphics[width=.75\textwidth]{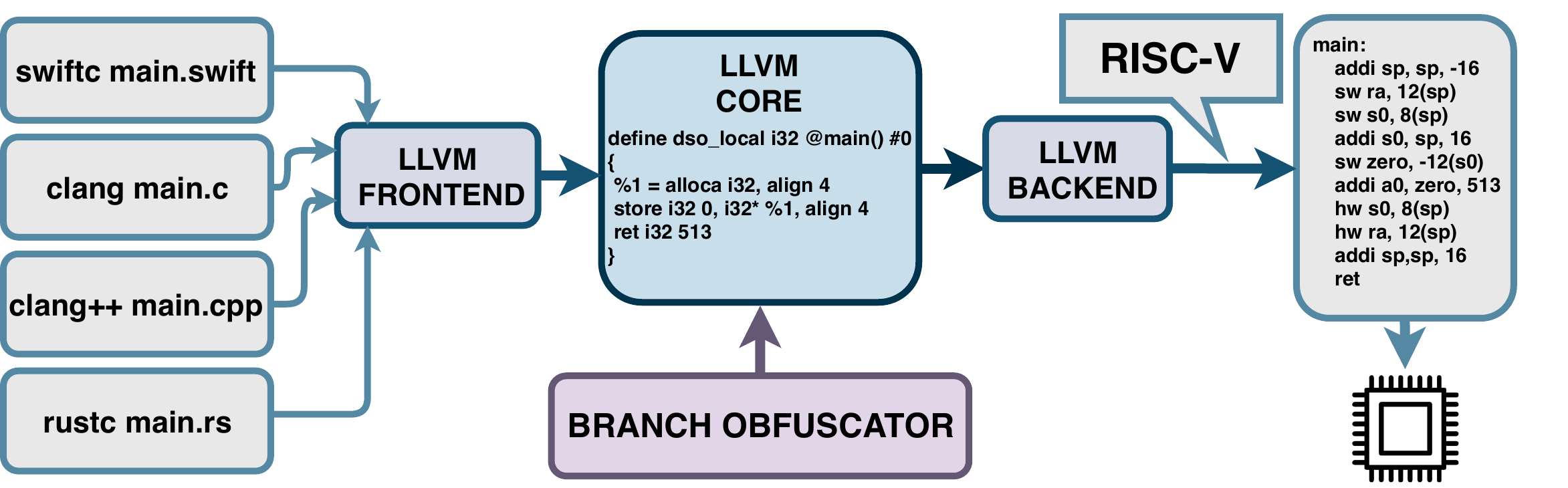}
  \caption{Place of the obfuscator in the LLVM compilation
    process. The obfuscation is done as the very last optimization.}
  \label{fig:obfuscator}
\end{figure*}

The obfuscator itself is implemented as an LLVM Pass
--- the same technique that LLVM compiler infrastructure uses
internally for its optimizations. The place of the obfuscator in the
LLVM compilation chain is given in Figure~\ref{fig:obfuscator}.  At
the beginning, a high-level programming language is translated to LLVM
intermediate representation (IR) using the corresponding compiler
frontend. After the IR is obtained, LLVM's optimizer performs the
optimizations in the specified order. The obfuscator takes control
over the process as the very last optimization and performs the
obfuscation immediately before the RISC-V backend. Since the branch
obfuscation is done at the level of LLVM IR, our technique supports
all the programming languages with compilers that target LLVM IR.\
This group of programming languages includes C, C++, Rust, Apple Swift
and others.

Another reason to build our obfuscation as an optimization on LLVM IR
is our need to have a stealthy obfuscation. In other words, we are
minimizing the number of properties that all obfuscated programs have
in common even before considering conditional branches. By doing so,
we aim at hindering adversarial queries of the form ``is this binary
even obfuscated or not''. A hypothetical obfuscation that would
perform on the resulting platform-specific assembly code would risk to
make the transformations that naturally cannot come from the given
compiler. In a realistic scenario, an attacker could know the compiler
that the victim software owner uses. From that information an attacker
could infer that the given binary cannot come from the known
compiler. For example, an attacker might consider the registers
utilization patterns in the obfuscated binary. The information on
register utilization patterns can be obtained using an advanced
reverse engineering tool as described in Section~\ref{sec:angr}.

%% file: latex_figures/transition_table.tex
\begin{table}[h!]
  \definecolor{navyblue}{rgb}{0.0, 0.0, 0.5}
  \definecolor{palecarmine}{rgb}{0.69, 0.25, 0.21}
  \footnotesize
  \begin{center}
    \caption{Basic table of translations used if the hash function
      returns a 1. \texttt{IF\_END} labels the basic block that
      is not executed when the branch with the condition from the
      first column is taken. \texttt{IF\_THEN} labels the basic block
      that is executed if the branch is taken.}
    \label{tab:transitions}
    \begin{tabular}{|l|p{28mm}|p{28mm}|}
      \hline
      \textbf{Condition} & \textbf{Original branch} & \textbf{Obfuscated branch} \\
      \hline
      \hline
      x1 == x2 & \textcolor{navyblue}{\textbf{bne x1, x2}}, .IF\_END \newline j .IF\_THEN & \textcolor{palecarmine}{\textbf{beq x1, x2}},
                                                                                            .IF\_END
                                                                                            \newline j
                                                                                            .IF\_THEN
      \\
      \hline
      x1 != x2 & \textcolor{navyblue}{\textbf{beq x1, x2}}, .IF\_END \newline j .IF\_THEN & \textcolor{palecarmine}{\textbf{bne x1, x2}},
                                                                                            .IF\_END
                                                                                            \newline
                                                                                            j
                                                                                            .IF\_THEN
      \\
      \hline
      x1 < x2 & \textcolor{navyblue}{\textbf{bge x1, x2}}, .IF\_END \newline  j .IF\_THEN & \textcolor{palecarmine}{\textbf{blt x1, x2}},
                                                                                            .IF\_END
                                                                                            \newline j
                                                                                            .IF\_THEN
      \\
      \hline
      x1 > x2 & \textcolor{navyblue}{\textbf{bge x2, x1}}, .IF\_END \newline j
                .IF\_THEN & \textcolor{palecarmine}{\textbf{blt x2, x1}}, .IF\_END
                            \newline j
                            .IF\_THEN
      \\
      \hline
      x1 <= x2 & \textcolor{navyblue}{\textbf{blt x2, x1}}, .IF\_END \newline j .IF\_THEN & \textcolor{palecarmine}{\textbf{bge x2, x1}}
                                                                                            .IF\_END
                                                                                            \newline
                                                                                            j
                                                                                            .IF\_THEN
      \\
      \hline
      x1 >= x2 & \textcolor{navyblue}{\textbf{blt x1, x2}} , .IF\_END \newline j .IF\_THEN & \textcolor{palecarmine}{\textbf{bge x1, x2}},
                                                                                             .IF\_END
                                                                                             \newline
                                                                                             j
                                                                                             .IF\_THEN
      \\
      \hline
    \end{tabular}
  \end{center}
\end{table}


%% file: deobfuscate.tex
\section{Deobfuscation Procedure} \label{lab:deobf}
\begin{figure*}[!ht]
    \centering
    \begin{minipage}{.48\textwidth}
        \centering
        \includegraphics[width=\columnwidth]{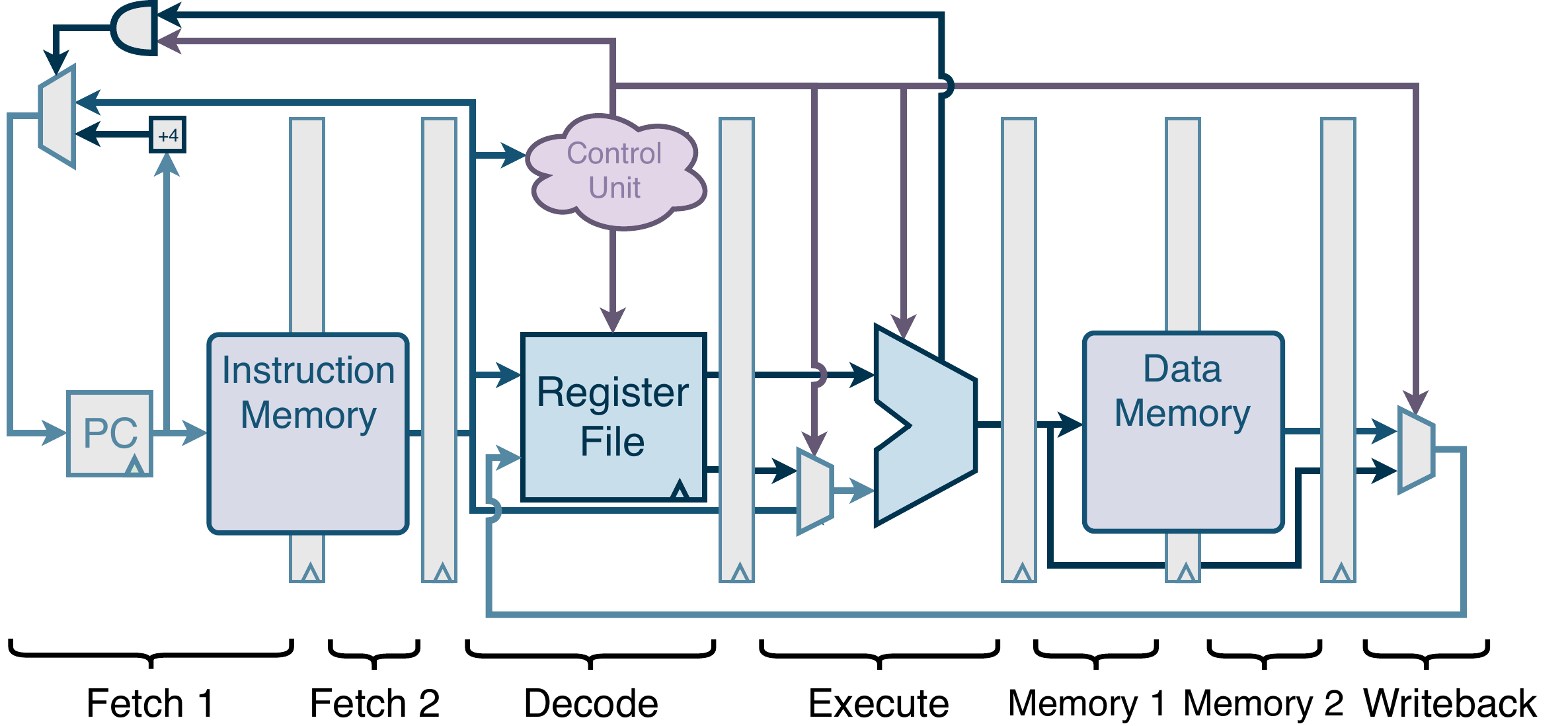}
        \caption{The original 7-stage RISC-V processor adopted from the BRISC-V platform~\cite{brisc_v}.}
        \label{fig:arch1}
    \end{minipage}%
    \hfill
    \begin{minipage}{.48\textwidth}
        \centering
        \includegraphics[width=\columnwidth]{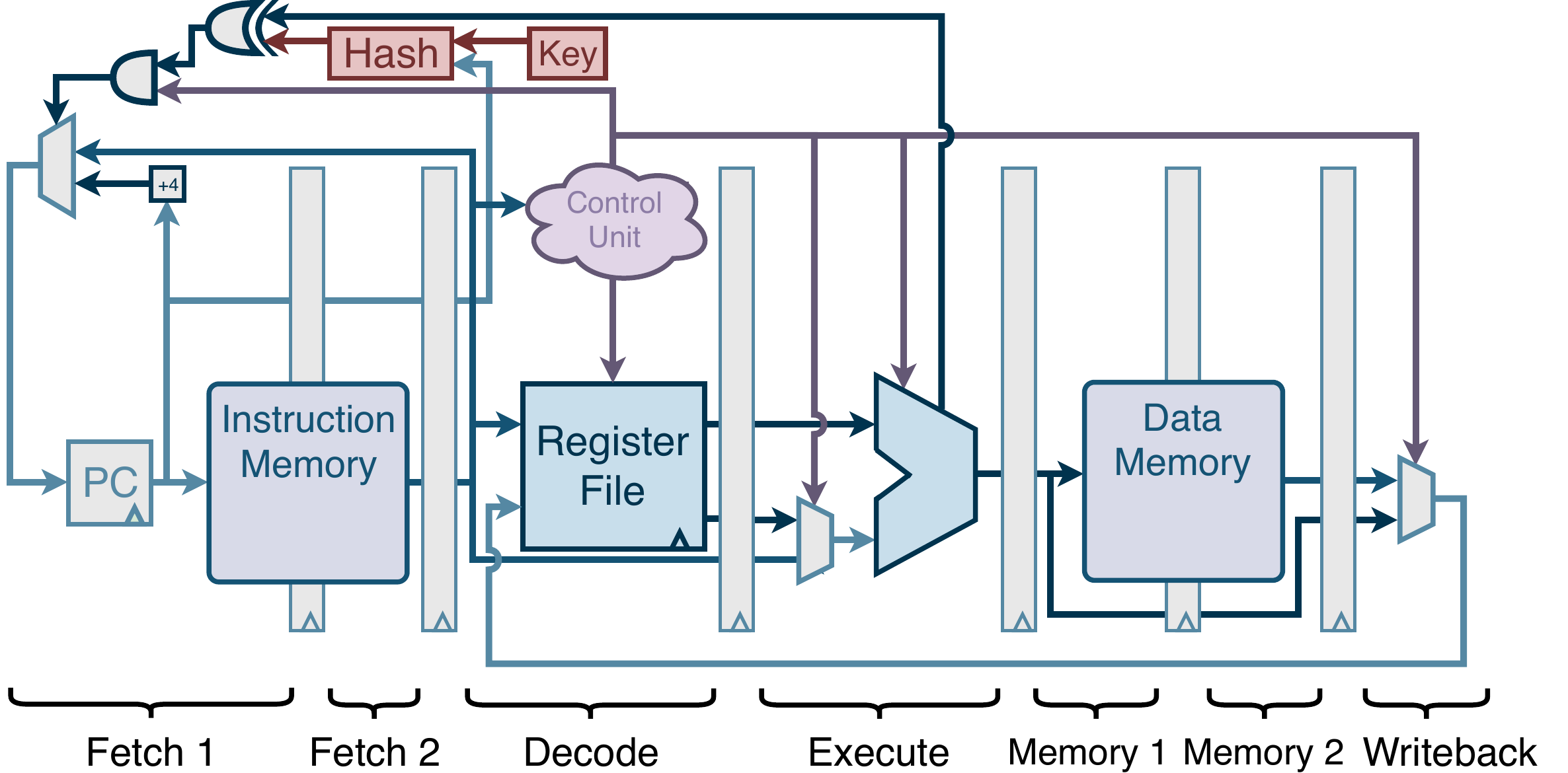}
        \caption{The 7-stage pipeline RISC-V processor modified to
        \texttt{XOR} branch values with hash function outputs.}
        \label{fig:arch2}
    \end{minipage}
    \caption{The baseline and the stalled-hash processor diagrams.}
\end{figure*}

The obfuscated program correctly executes only on a trusted RISC-V
core designed to support deobfuscation and supplied with the secret
key. We outline four designs here: the baseline, stalled-hash,
cached-hash and the mask-based design.

\noindent\textbf{Baseline design:} the baseline design is a 7-stage
RISC-V CPU without any hardware modifications enabling obfuscation~\cite{brisc_v}.
The processor uses synchronous block RAM (BRAM) for instruction and
data memories, which requires 2 extra stages over a typical 5-stage processor.
A simplified processor architecture is illustrated in
Figure~\ref{fig:arch1}.

\noindent\textbf{Stalled-hash design:} here, the baseline CPU is
equipped with a hardware hash function. When a branch instruction is in
the decode stage, the hash function is fed with the branch instruction
address and the program key. When the branch instruction reaches the
execute stage, all the stages up to and including the execute stage
are stalled until the hash function produces an output. Once a (single
bit) output is produced, that value is \texttt{XOR}-ed with the branch
signal. This way, branches that would be taken may not be, and
vice-versa. For the hash function, we use a parametrizable
Linear Feedback Shift Register (LFSR) with
$n$ registers, and let the LFSR run for $k, k > n$
cycles. A cryptographically secure pseudo-random number generator may provide
higher security at the cost of an increased latency, causing the processor
to stall more. As only a single hash function is calculated at any given time,
pipelining the hash function brings no benefit. Additionally, since the
output of the hash is pseudo-random, a branch predictor may at best have
a 50\% chance of guessing the branch result.
The modified architecture is illustrated in Figure~\ref{fig:arch2}.

\noindent\textbf{Cached-hash design:} since the hash function output is
only dependent on the address of each branch, the hash of a given branch is
constant. This allows us to cache the hashes of previously evaluated branches.
In this design, we add a cache to the stalled-hash design.
When a branch is in the decode stage,
the architecture starts calculating the hash function and
in parallel checks the cache for whether that branch's hash has previously been
calculated. If not, when the hash function finishes, it both feeds the
value to the XOR gate, and saves the result to the cache. If the value
is found in the cache, it is sent to the XOR, just in time as the
branch enters the execute stage, causing no stall. In our experiments we used a simple
256-line, one branch per line, direct-mapped cache.

\noindent\textbf{Mask-based design:} here, the baseline CPU is modified
so that the instruction memory is extended with a single `mask' bit.
The mask bit specifies whether a branch should be reversed or not.
The mask bit follows the instruction through the stages, and is consumed
in the execute stage if the instruction is a branch.
Having an independent mask bit per each branch removes the possibility of
an attacker predicting future branches based on the existing ones. However,
widening the instruction word width complicates the design of L2 caches
and memory controllers. Furthermore, depending on the attack model,
the masks may need to be kept encrypted in memory and decrypted on-the-fly.

%% file: deobfuscation_results.tex
\section{Reverse Engineering of Obfuscated Code} \label{sec:angr}
Offensive binary analysis is a mixture of static and dynamic
techniques that discover crashing inputs. Besides the trivial usage in
compromising system's availability, this family of techniques can be
used in attack surface exploration. Some of the discovered crashes
(overwritten function pointers, buffer overflows, etc.) can be used by
attackers to take control of a program's execution. One of the
possible attacks applicable to the results of offensive binary
analysis is return-oriented programming~\cite{roemer2012return}. This
particular technique draws much attention because it does not
require any code injection. Instead, it constructs malicious actions
from the instructions already present in the address space of the
executed program.

To show the behavior of the obfuscated binary under offensive analysis
we use Angr~\cite{Shoshitaishvili_2016}. Its core binary analysis
features rely on concolic execution. This technique combines symbolic
and concrete execution to construct the inputs that spot the
vulnerabilities and eventually exploit them. Further, we briefly
discuss the structure and function of Angr without claiming credit for
its design or implementation.

\begin{figure*}[!h]
  \centering
  \includegraphics[width=0.75\textwidth]{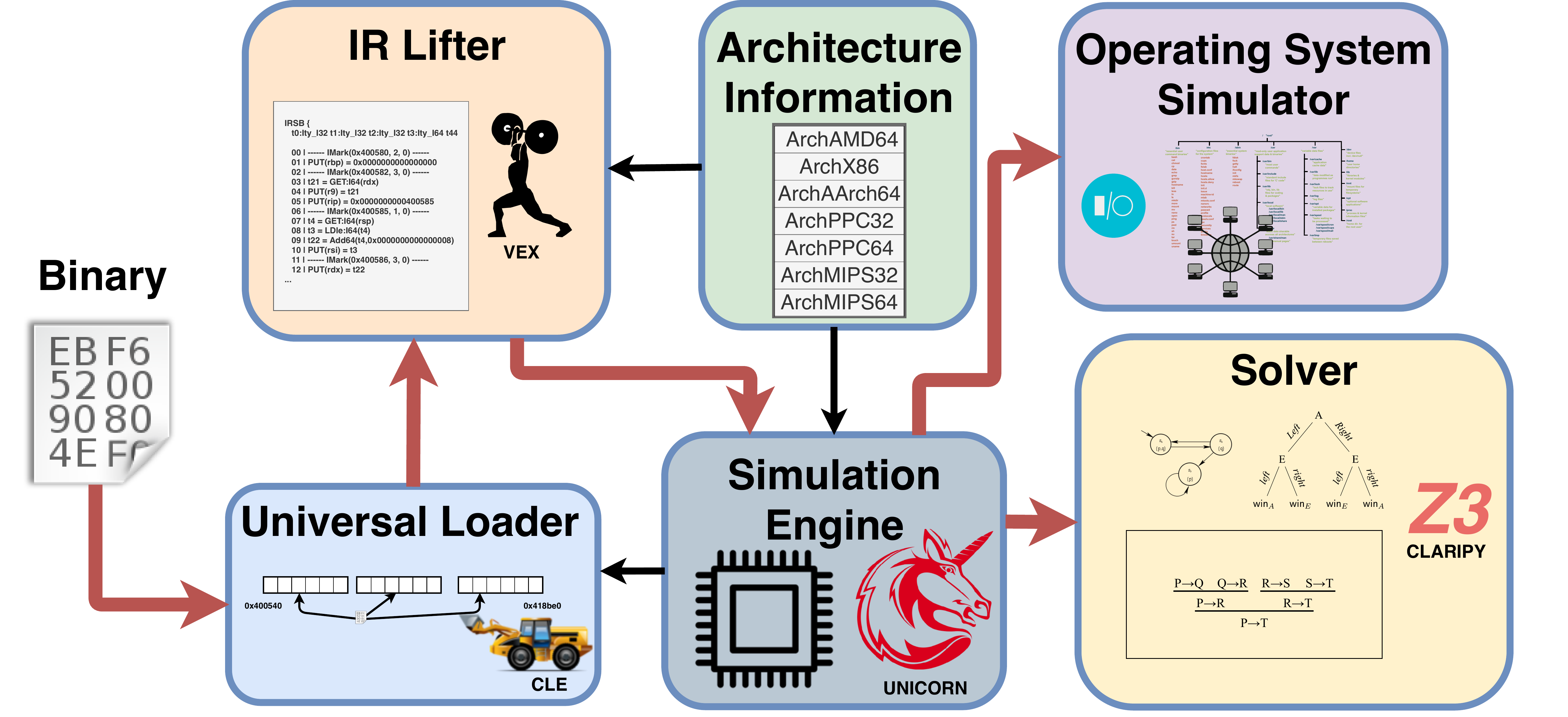}
  \caption{The structure of Angr.}
  \label{fig:angr}
\end{figure*}

Angr can analyze the binaries packed without the symbol table and
relocation information --- stripped binaries. When the binary is
given, the analysis process flows as shown in Figure~\ref{fig:angr}.
\begin{description}[leftmargin=0pt, labelindent=0pt]
\item[Binary loading.] A universal loader creates an address space in
  which it organizes all the needed binaries. The abstraction of the
  address space is returned to the system. Besides ELF binary format,
  Angr also supports others such as Microsoft Windows PE.\ When the
  binary format is recognized, its header is used to fetch the
  architecture information.
\item[Lifting.] Angr can analyze binaries targeting multiple architectures. From the
  functional perspective, certain architectures are similar to each
  other enough to be handled by a single abstraction. General analysis
  algorithms are then applied to the abstraction instead of each
  platform separately. For this purpose, Angr reuses the Valgrind
  VEX~\cite{valgrind} intermediate representation. Roughly speaking,
  VEX abstraction consists of
  \begin{enumerate*}[label=\arabic*)]
  \item establishing a unique register naming for all the platforms,
  \item removing the differences in memory accesses and segmentation, and
  \item unfolding the instructions with side effects to make them
    transparent.
  \end{enumerate*}
\item[Simulation.] In accordance with architecture information, the
  tool chooses the adequate simulation engine. Such an engine uses
  basic blocks as the portions of program's execution. It further
  interprets the portions starting with an input state that comprises
  of the register snapshot, memory etc. The results of the simulation
  step are all the possible successor states. While the simulation
  explores the branches, it collects the branch
  conditions. Subsequently, each resulting successor state contains
  its reachability condition. This is the prerequisite for the cyber
  reasoning features in Angr. However, when the tool explores our
  obfuscated binary, it collects the branch conditions whose
  correctness depends on whether obfuscator performed the reversal or
  not. If the collected branch condition remained intact during
  obfuscation, the simulation engine will discover it
  correctly. Otherwise, if the simulation engine traverses at least
  one reversed branch on its way to the successor state, its
  reachability condition is rendered incorrect. Under the premise of
  safe key distribution, by no means can the simulation discover the
  correct branch conditions for all the branches in a non-trivial
  program.

\item[Constraint solving.] All the simulation steps that the simulator
  undertakes may be performed on concrete or symbolic values. Symbolic
  values can be complex expressions and the operations on them result
  in the new symbolic expressions. For example, if the value of the
  program counter depends on the input of the program, all the
  transitions will be represented by the symbolic expressions. The
  core of the constraint solver deployed in Angr leans on
  Satisfiability Modulo Theories~\cite{barrett2018satisfiability}
  carried by Microsoft's Z3 solver~\cite{de2008z3}. This allows
  an attacker to efficiently explore the inputs that lead to
  vulnerabilities. However, our obfuscator eliminates the threat
  imposed thereof since it makes the simulator construct the wrong
  constraints. A quick and correct solution to the wrong constraints
  does not equip the attacker with a functional
  vulnerability-discovering input.
\item[Operating system simulation.] To simulate program's interaction
  with the operating system, Angr deploys the internal implementation
  of system calls. It supports multiple operating system kernels
  including Linux. The simulated system calls take effect on the
  simulated states that normally contain symbolic values.
\end{description}


%% file: evaluate.tex
\section{Evaluation}\label{sec:eval}
To evaluate the performance of Drndalo we set up a series of
experiments. The experiments are divided in two groups based on the
implementation of the deobfuscation phase. First, we evaluate the
performance and assess the security offered by experimental
in-software deobfuscation procedures. Second, we measure the
performance of various in-hardware deobfuscation implementations and
discuss the security properties of each.

\subsection{In-software Deobfuscation}
Due to our attack model, a deobfuscation procedure must be capable of
operating on the binary without debug symbols or other additional
information --- stripped binaries. To precisely define the space of
possible designs, we impose three strict requirements to all
in-software deobfuscation procedures:
\begin{enumerate*}[label=\arabic*)]
\item knowing the correct key, a procedure must completely recreate
  the original program semantics,
\item deobfuscation must maximally avoid storing the plain version of
  the critical program so as to minimize the attack surface, and
\item in doing so, the procedure does not require any hardware
  modifications and runs on commodity hardware.
\end{enumerate*}

\begin{figure}
  \centering
   \includegraphics[width=\columnwidth]{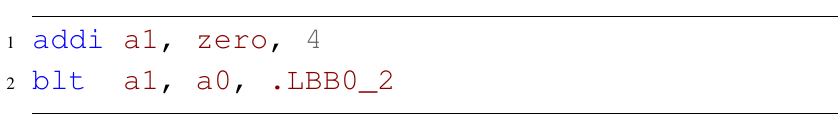}
  \caption{Original assembly before runtime obfuscation.}
  \label{lst:runtime_plain}
\end{figure}

\begin{figure}
  \centering
  \includegraphics[width=\columnwidth]{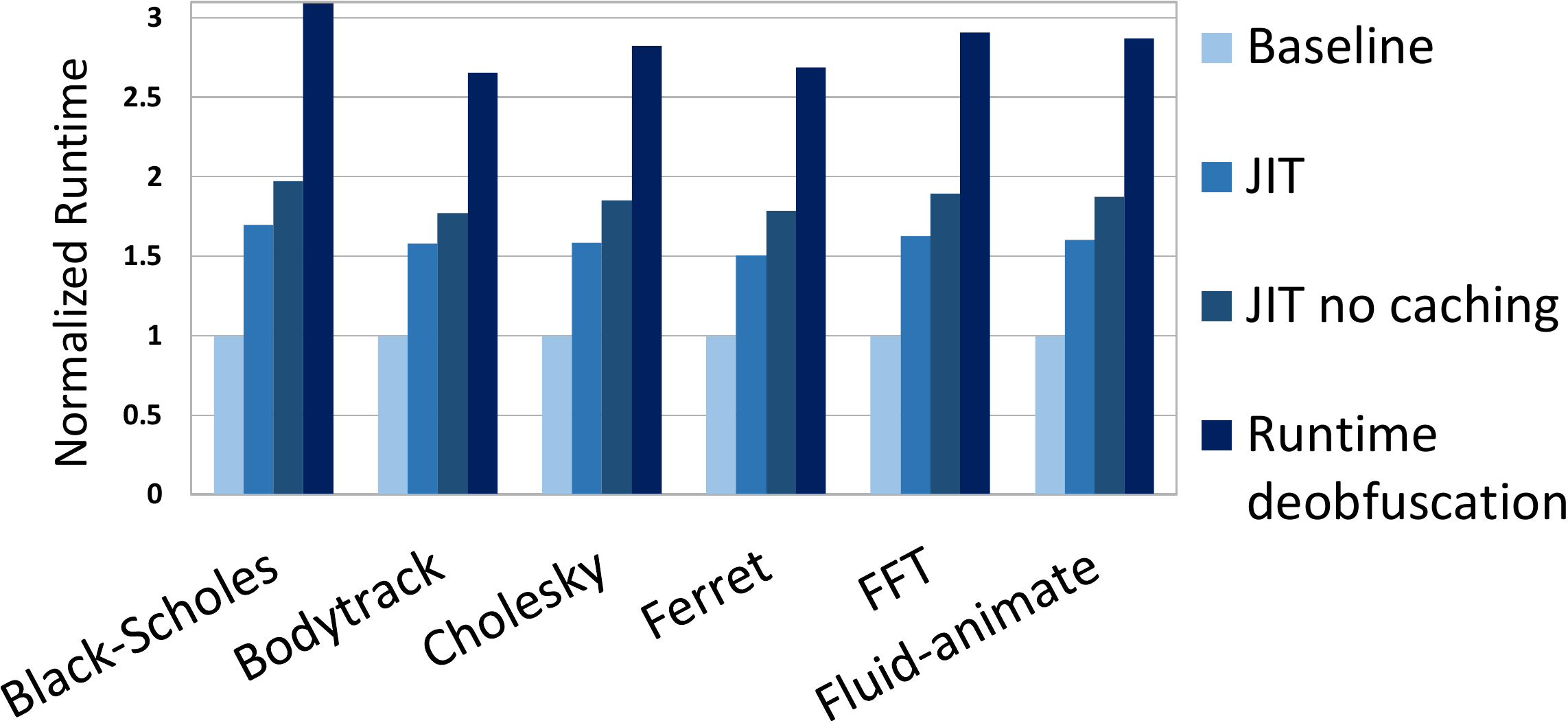}
  \caption{In-software deobfuscation on selected PARSEC benchmarks.}
  \label{fig:in-software}
\end{figure}

\begin{figure}
  \centering
  \definecolor{bg2}{rgb}{0.90,0.90,0.90}
	\includegraphics[width=\columnwidth]{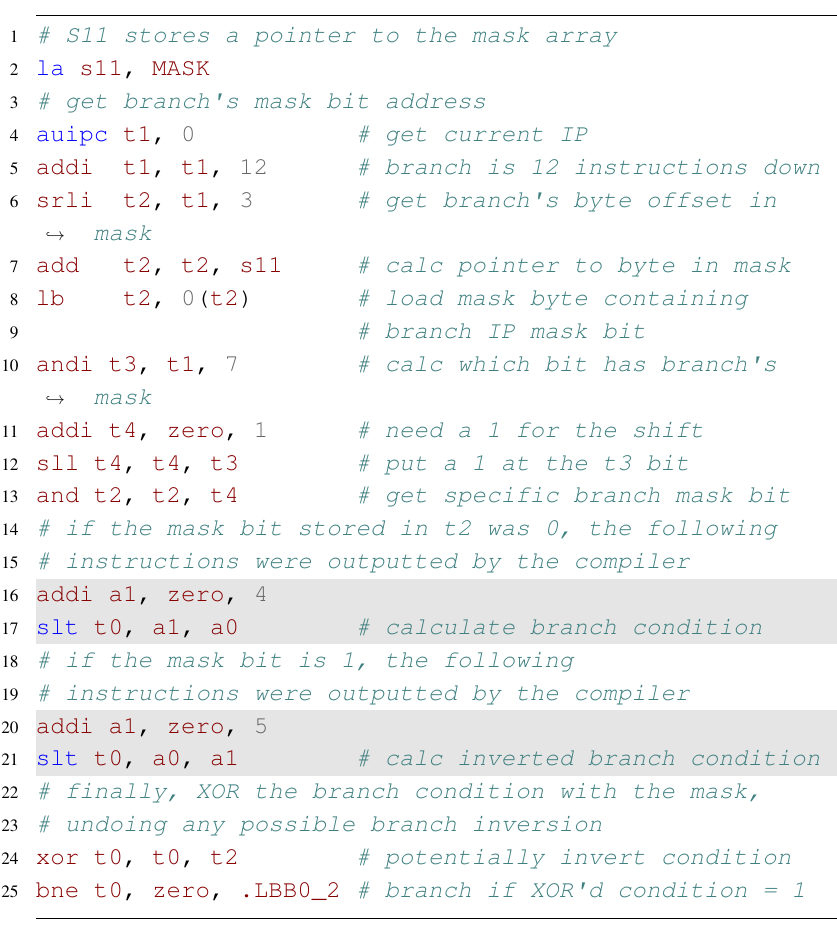}
  \caption{Assembly after runtime obfuscation.}
  \label{lst:runtime_obf}
\end{figure}

According to the mechanics used for restoring the original program
from the obfuscated code and the secret key, we outline two families
of in-software deobfuscation procedures:

\begin{description}[leftmargin=0pt, labelindent=0pt]
\item[JIT-based deobfuscation.] This approach is based on traditional
  just-in-time compilation techniques. The JIT compiler implements the
  inverse transformation of the Table~\ref{tab:transitions} and
  incurs no other transformation. For example, given the code from
  the Figure~\ref{lst:obf}, the compiler produces the code from the
  Figure ~\ref{lst:plain}. The compiled instructions are stored and
  only need to be deobfuscated once. Hence, the JIT penalty is
  proportional to code size, and not the program runtime.

  Our experimental in-software deobfuscation procedure is implemented in
  Intel Pin dynamic binary instrumentation
  framework~\cite{luk2005pin}. This implementation of JIT compiler
  operates on the basis of basic blocks and inspects all the conditional
  branches. The basic steps of the JIT compiler are as follows:
  \begin{enumerate*}[label=\arabic*)]
  \item load the secret key,
  \item instrument all the conditional branches,
  \item apply the hash function to the conditional branch under
    consideration,
  \item emit the branch instruction with the appropriate reversed
    condition (Table~\ref{tab:transitions} obfuscated branch to original
    branch),
  \item execute the next basic block.
  \end{enumerate*}
  The simplest possible construct that can serve a similar purpose as a
  hash function is a bit mask of inversions generated by the obfuscator
  at compile time. In the absence of the key, we assume that the bit
  mask is kept secret. Since each branch has the same probability of
  being inverted, we assume that 50\% of the inversion bits are set. Our
  experimental deobfuscation procedure caches the branches to assure
  that no branch inversion is checked twice. Thus, our procedure needs
  to:
  \begin{enumerate*}[label=\arabic*), ref=\arabic*]
  \item load the corresponding inversion bit from the mask,
  \item execute a compare instruction on the mask bit, and
  \item \label{item:inverse} execute the logic of reverting, if needed.
  \end{enumerate*}
  There are many different ways for the JIT-compiler to perform the
  step~\ref{item:inverse} once it knows that it is needed. However, we
  will make an optimistic assumption that the logic of inversion will
  not take more than 10 instructions per execution. The optimistic
  estimates for both cached and non-cached JIT-compiler performance
  overheads are given in
  Figure~\ref{fig:in-software}. The main difference
    between JIT with and without caching is that when caching is
    disabled, the branch deobfuscation procedure is repeated for the
    repeating branches (e.g., branches in the loops). When caching is
    enabled, the result of the deobfuscation procedure is reused.
\item[Runtime deobfuscation.] Another in-software deobfuscation technique
  is to force the obfuscator to emit the code that contains the branch
  calculation in runtime. Here the code is obfuscated so as to load
  from the mask or calculate using the hash function whether each
  branch should be reverted at runtime. Given the original assembly in
  Figure~\ref{lst:runtime_plain}, and opting to use the bit mask in
  obfuscation, the obfuscator outputs assembly as in
  Figure~\ref{lst:runtime_obf}. When run, this code will find the
  appropriate mask bit, calculate the branch condition, and branch
  depending on the \texttt{XOR}-ed value of the two bits. Unlike in
  the case of the JIT implementation, the runtime penalty is
  proportional to the number of branches executed, and not code size
  (as in the case of a JIT implementation).
  As shown in Figure~\ref{lst:runtime_obf}, runtime deobfuscation
  approach adds several instructions per each branch, leading to
  significant code bloat and a longer
  runtime. Figure~\ref{fig:in-software} shows the overhead of the
  runtime obfuscation being significantly higher than its JIT-based
  counterpart. Hence, we opt for JIT-based deobfuscation as a more
  efficient in-software solution.
\end{description}

\subsection{In-hardware Deobfuscation}
\begin{figure*}[!h]
  \centering
  \includegraphics[width=.95\textwidth]{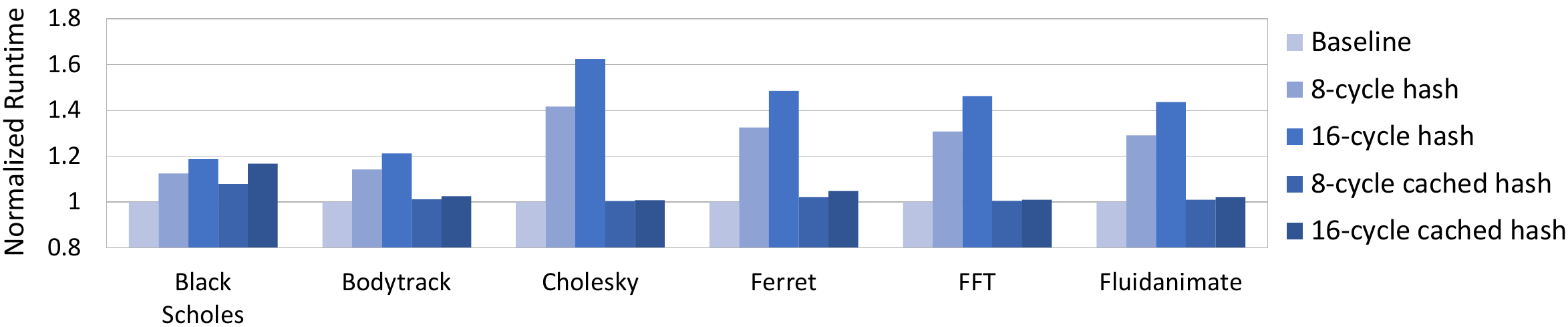}
  \caption{The performance of the five architectures on 6 PARSEC tasks, normalized for the baseline processor.}
  \label{fig:eval}
\end{figure*}

We simulate the four architectures listed in Section~\ref{lab:deobf}. The
baseline and the mask-based architectures have approximately the same performance,
so we omit the second one. Further, we test two different hash-based
architectures with an 8 and a 16-cycle hash function. In
Figure~\ref{fig:eval}, we see the performance of different
architectures on 6 different PARSEC tasks. Notice that for the
16-cycle hash, the processor can slow down as much as 60\%. However,
adding a (256-line, single branch hash per line) direct-mapped cache
removes the majority of the performance overhead. As opposed to the
in-software implementations from Figure~\ref{fig:in-software},
the in-hardware techniques achieve a negligible average runtime
overhead of around 5\%.

\subsection{Hardware utilization}
\begin{figure}[b]
  \centering
  \includegraphics[width=\columnwidth]{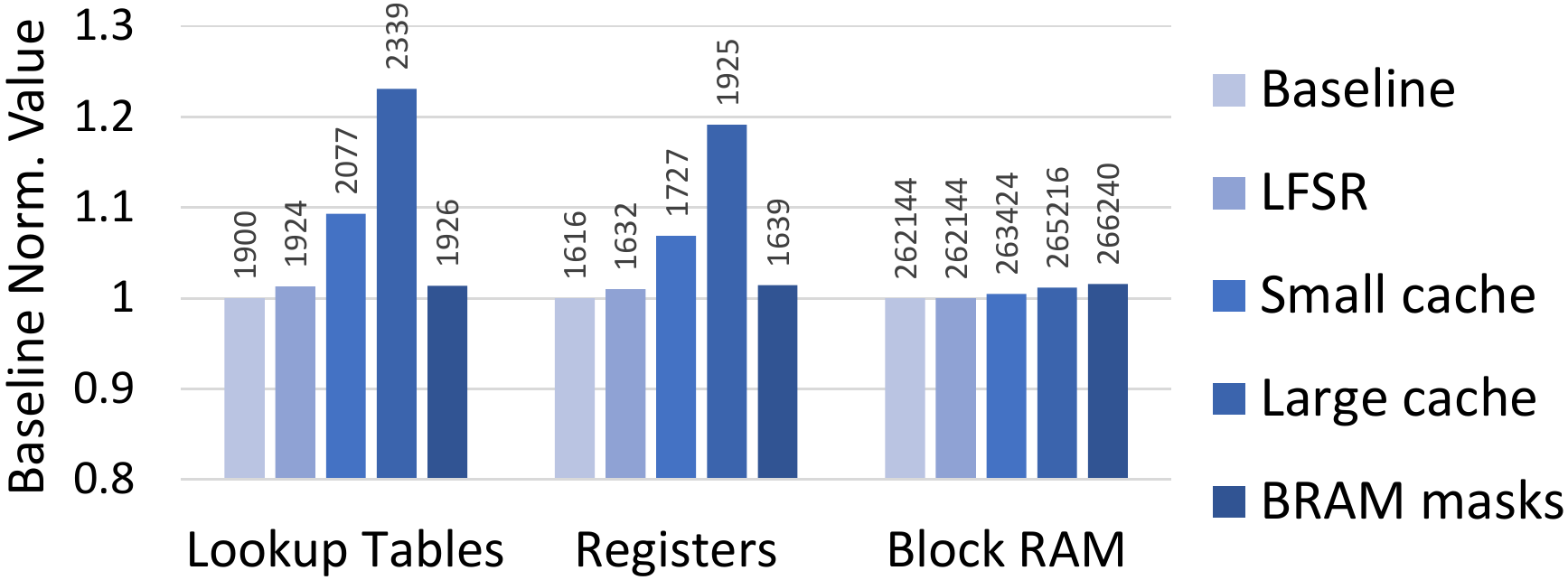}
  \caption{Lookup table (LUT), register, and BRAM usage. The LFSR, small cache, and large cache designs all use a 16 bit LFSR. The small and large cache designs have a 256 and 1024-line direct mapped-cache, respectively.}
  \label{fig:area}
\end{figure}

We implement the  designs from Section~\ref{lab:deobf} as modifications to a baseline
RISC-V CPU published in~\cite{brisc_v}. We synthesize 5 different designs using
Altera Quartus:
\begin{description}[leftmargin=0pt, labelindent=0pt]
\item [Baseline:] a baseline 7-stage in-order RISC-V CPU, with BRAM-based instruction and
  data caches. The design is synthesized with 16KB of instruction and data
  memory, hence the BRAM utilization of the baseline design in
  Figure~\ref{fig:area}.
\item[Stalled-hash:] a modified design with a linear feedback shift register (LFSR)-based
  hash function that stalls the design for a parametrized number of cycles
  on every branch instruction. We choose an LFSR over a more cryptographically
  secure random number generator due to it's small size and efficiency.
\item[Cached-hash:] an extended LFSR-based design with a cache used for storing the LFSR
  outputs of previously seen branches. Each cache stores a single bit hash value
  corresponding to an instruction address. The cache is accessed in the decode stage,
  and returns a match by the execute stage. If a match is found, the design does
  not stall at all. We test two configurations: a 256
  and a 1024-line direct-mapped cache. In the current implementation, the cache
  is synthesized using registers, but can be stored in BRAM too.
\item[Mask-based:] a mask-based design, with an additional BRAM-based cache storing one 1-bit
  hash value per each instruction. This value is only used in branch instructions, and is otherwise
  ignored.
\end{description}

The LUT, register, and block RAM counts of synthesized designs are
shown in Figure~\ref{fig:area}. Note that we have implemented an separate hash cache from the instruction
cache. The majority of memory used for this cache is spent on tags, not hash values. In future work
we plan to explore an implementation that extends the L1 cache with the extra hash bit, removing
the need for storing the tags twice.


%% file: security_evaluation.tex
\section{Security Evaluation}\label{sec:sec_eval}

The security of our obufscation method relies on
\begin{enumerate*}[label=(\arabic*)]
\item the security of the hash function, and
\item the security of the branch obfuscator module.
\end{enumerate*}
Assuming that the hash function is cryptographically safe, in this
section we discuss the security of the obfuscation method.

We empirically evaluate the security of our obufscation method by
attempting to deobfuscate binaries using a classifier trained on
both obfuscated and plain binaries from a dataset of known programs.
This type of attack is plausible since the obfuscation method is
published and the attacker can freely create a dataset of plain and
obfuscated binaries. Hence, an attacker may be able to use an in-house
trained classifier to reconstruct an original control flow of an
obfuscated binary whose source they do not possess.

The classifier training pipeline is composed of:
\begin{enumerate}
\item \textbf{Dataset collection:} to create the dataset, we compile
  and obfuscate 98 programs from the LLVM test suite single source
  benchmarks \footnote{github.com/llvm-mirror/test-suite/tree/master/SingleSource}.
  \item \textbf{Feature extraction:} since the obfuscation is
  applied after the compiler front-end, both the plain
  and the obfuscated binaries will have the same topology,
  i.e., the control flow graph of a binary before and after
  obfuscation remains the same. This allows us to simply
  compare modified (obfuscated) and unmodified branching
  basic blocks (BBBl). We use Angr to in-parallel parse plain
  and obfuscated binaries, detect which BBBl's have been
  obfuscated, and store modified/unmodified BBBl's into
  separate categories. We store several features of BBBl's,
  shown in Table~\ref{table:features}. As BBBl's have
  different numbers of instructions, we only record a window
  of the last $I$ instructions before (and including) the
  branch.
  \input{./latex_figures/feature_table.tex}
\item \textbf{Preprocessing:} we use cross-validation and create
  two datasets: a training and a test one. To prevent the
  classifier from learning binary-specific features, the test
  set is created with \textbf{separate} binaries from the
  training set (i.e., not just separate basic blocks).
  Since only the obfuscated binaries have obfuscated BBBl's,
  and only 50\% of those BBBl's will be obfuscated, the ratio
  of plain to obfuscated BBBl's is 3/1. To balance the two
  classes, we normalize the training set to only include one
  third of plain branches. Next, since opcodes and register
  indexes are integers and it is meaningless to compare
  instructions by their opcode values, we convert all integers
  to one-hot vectors.
\item \textbf{Training:} we test several different
  classifiers. Each classifier is fed with a number of
  concatenated one-hot vectors, and is trained to predict the
  $OBF$ bit, specifying whether the branch has been obfuscated.
\end{enumerate}

\subsection{Classifier Results}
We test several classifiers, including: logistic regression, decision
trees, random forest classifiers, and multi-layer perceptrons. All of
the classifiers seem to achieve very similar results, with an average
classification accuracy of 63\%. We give the confusion matrices of
different classifiers in Figure~\ref{table:confusion}.
\input{./latex_figures/confusion_matrices.tex}

\begin{figure}[h!]
  \centering
  \includegraphics[width=1.0\columnwidth]{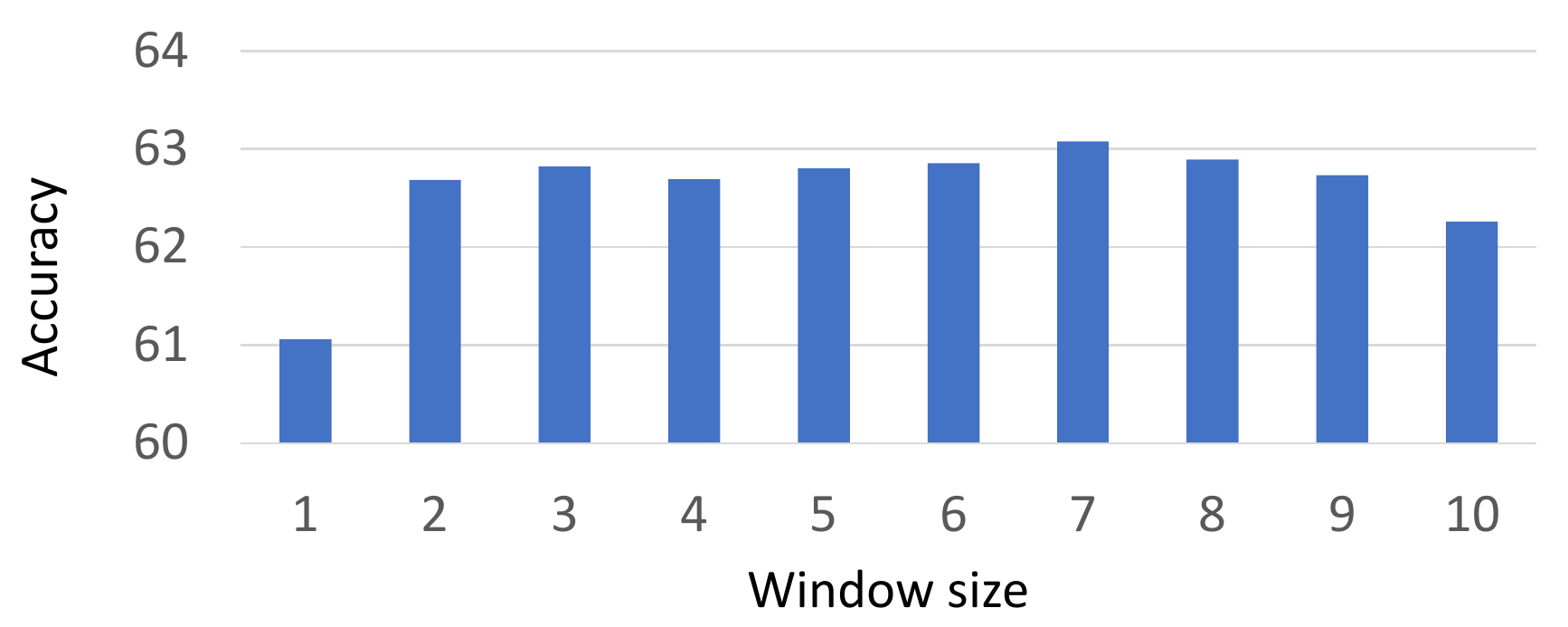}
  \caption{The average accuracy of the classifier over 10 experiments with respect to the
    instruction window size.}
  \label{fig:window_size}
\end{figure}

In order to test whether the classifier only learns some statistical
information (i.e., that certain branch instructions are more common in
obfuscated binaries compared to plain ones) or if the classifiers
found a pattern in the produced instructions, we vary the size of the
instruction window $I$. In Figure~\ref{fig:window_size}, we show the
accuracy of classifiers trained on features with different windows
sizes.

\subsection{Analysis of Classifier Results}

As our obfuscation method \texttt{XOR}s branch predicates with random
bits, we would expect that the outputted values are indistinguishable
from randomness. Hence, any classifier should be unable to predict which
branches were obfuscated with an accuracy higher than 50\%, assuming that
the number of obfuscated and plain branches is equal.
However, Table~\ref{table:confusion} shows that all classifiers
have some success in predicting which branch is obfuscated, with an
average prediction accuracy of 63\%.
Interestingly, the majority of classifiers seem to be unable to
distinguish whether a truly obfuscated branch is obfuscated or plain,
but can consistently recognize truly plain branches as plain.

To understand why classifiers are able to reach accuracies higher than
50\%, we train classifiers with varying window sizes, as seen in
Figure~\ref{fig:window_size}. We see that the majority of the accuracy
gains is achieved with a window size of 1, and that adding more
instructions does not significantly improve performance. With only access
to the branch instruction opcode $OP_{\ 0}$, branch registers $R1_0, \; R2_0, \; RW_0$,
and the $BR_{up}$ bit specifying whether the branch instruction is
branching to an address above or below it, the classifier is able to get
an 11 percentage point increase over random guessing. Hence, the classifier
is not learning any specific pattern of instructions that may reveal
information about the original LLVM IR or the source code, but is instead
relying on the distribution of different instructions in the dataset of
plain and obfuscated binaries. As a simple example, consider a non-obfuscating
compiler that outputs certain branch instructions, e.g., \texttt{BLE} and
\texttt{BLT} with a 60\% and 40\% probability, respectively.
The obfuscating compiler may shift these probabilities to 50\% and 50\%.
Hence, whenever the attacker decodes a \texttt{BLT} instruction, they
can claim that the branch is obfuscated with ~60\% probability.

\begin{figure}[h!]
  \centering
  \includegraphics[width=0.8\columnwidth]{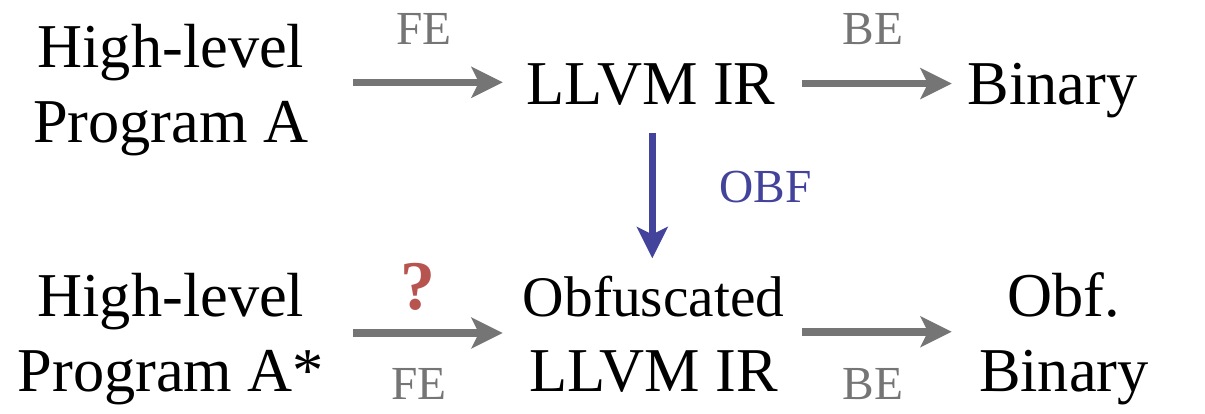}
  \caption{Compilation flow of a high-level language program $A$, with
  and without obfuscation. We ask whether there exists a program $A*$ such
  that the LLVM front-end produces the obfuscated LLVM IR. FE and BE stand
  for compiler front-end and back-end, and OBF stands for the Drndalo compiler
  extension.}
  \label{fig:flow}
\end{figure}

One way to present this is shown in Figure~\ref{fig:flow}. Given an
original high-level language program $A$, the LLVM front-end produces
LLVM IR, and the back-end produces the binary. If Drndalo is used to
obfuscate the IR, the obfuscated IR is used to produce the
binary. However, some obfuscated programs in the form of LLVM IR are
unlikely to originate directly from a program expressed in a high-level
language such as C. The classifier recognizes such unlikely
program constructs and classifies them as obfuscated because they
rarely appear in plain binaries.

\begin{figure}[h]
  \centering
  \includegraphics[width=\columnwidth]{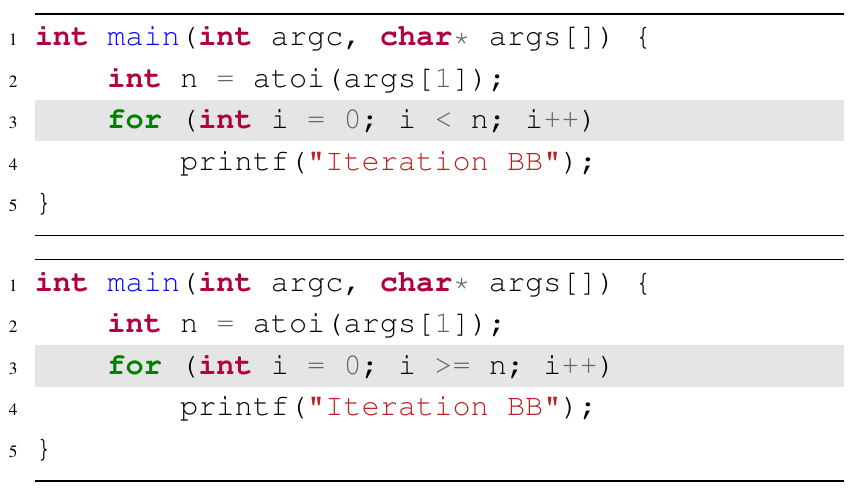}
  \caption{A ``for'' before obfuscation (top) and the program
    corresponding to its obfuscated version when the branch is
    inverted (bottom).}
  \label{lst:obf_for}
\end{figure}

To illustrate a scenario in which an unlikely LLVM IR is produced by
the obfuscator, consider the top program of Figure~\ref{lst:obf_for}
and assume that obfuscator inverts the loop's branch. For the sake of clarity
we will analyze the obfuscated program in its lifted version --- the
version in the C high-level language. If the lifted obfuscated program in the
bottom of Figure~\ref{lst:obf_for} is now analyzed, it can
be assessed as highly suspicious. Specifically, we see that the
\texttt{i} starts with zero, and increases as long as it is larger than \texttt{n}.
As the  variable \texttt{n} comes from a program input, we can detect two
possible outcomes of this ``for'' loop: either the input \texttt{n}
is zero or less and the loop executes infinitely many times, or
\texttt{n} is larger than zero and the loop never executes.
Thus, we can infer that this loop is either infinite or never executes.
As such a loop is an uncommon construct, we can infer that the loop is obfuscated.
The classifier can pick up such common and uncommon constructs and make
predictions based on them.

The example in Figure~\ref{lst:obf_for} points out a weakness of
Drndalo method that leaps beyond the scope of this work. The weakness
roots in the fact that not all valid semantic constructs in high
level programming languages are equally ubiquitous.


%% file: latex_figures/feature_table.tex
\begin{table}[h!]
\centering
\begin{tabular}{|c | c | c |} 
 \hline
 \textbf{Feature} & \textbf{Type} & \textbf{Description} \\ [0.5ex] 
 \hline\hline
 $\mathbf{OP_i   }$ & Int  & $i$-th instruction's opcode                     \\ \hline
 $\mathbf{R1_i   }$ & Int  & $i$-th instruction's 1st read register index  \\ \hline
 $\mathbf{R2_i   }$ & Int  & $i$-th instruction's 2nd read register index \\ \hline
 $\mathbf{RW_i   }$ & Int  & $i$-th instruction's write register index       \\ \hline
 $\mathbf{BR_{up}}$ & Bool & Is branching to a lower address?                \\ \hline \hline
 $\mathbf{OBF    }$ & Bool & Is branch obfuscated?                           \\ \hline
\end{tabular}
\caption{Features recorded for every branching basic block. For a 
window of size $I$, $i$ is the index of the instruction before the branch, $i \in \{0, 1, ..., I-1\}$}
\label{table:features}
\end{table}

%% file: latex_figures/confusion_matrices.tex
\begin{figure}[!ht]
   \subfloat[Decision tree, Acc: 62.8\% ]{%
       \begin{tabular*}{4cm}{c|cc}
               & PP  & PO  \\ \cline{1-3}
            TP & 544 & 138 \\
            TO & 369 & 313 \\
        \end{tabular*}
   }
   \hfill
   \subfloat[Random forest, Acc: 63.8\% ]{%
       \begin{tabular*}{4cm}{c|cc}
               & PP  & PO  \\ \cline{1-3}
            TP & 539 & 143 \\
            TO & 351 & 331 \\
        \end{tabular*}
   }
   \hfill\\
   \subfloat[Logistic regression, Acc: 63.4\% ]{%
       \begin{tabular*}{4cm}{c|cc}
               & PP  & PO  \\ \cline{1-3}
            TP & 533 & 148 \\
            TO & 350 & 331 \\
        \end{tabular*}
   }
   \hfill
   \subfloat[Multi-layer perceptron, Acc: 62.0\% ]{%
       \begin{tabular*}{4cm}{c|cc}
               & PP  & PO  \\ \cline{1-3}
            TP & 519 & 158 \\
            TO & 356 & 321 \\
        \end{tabular*}
   }
   \caption{Confusion matrices of 4 classifiers. TP, TO, PP, and PO labels stand for "true plain", "true obfuscated", 
   "predicted plain" and "predicted obfuscated".}
   \label{table:confusion}
\end{figure}

%% file: related_work.tex
\section{Related Work}\label{sec:rel_work}

Program obfuscation is a wide research subject and has historically
been approached from significantly different viewpoints. Some of the
research branches that tackle obfuscation are:
\begin{enumerate*}[label=\arabic*)]
\item malware packing~\cite{sharif2008,schrittwieser2013},
\item intellectual property
  protection~\cite{wermke2018,laszlo2009,Schrittwieser_2016,
    Zhuang_2004}, and
\item resistance to disassembly and reverse
  engineering~\cite{linn2003,Xu_2018}.
\end{enumerate*}
Normally, the approaches in literature serve one of the purposes
extremely well while fall short to protect against other
objectives. For example, the approach of Cousins et
al.\cite{Cousins_2018} obfuscates programs in a way that makes it
completely unintelligible for an attacker. Consequently, a program
obfuscated in such a way is virtually resistant to IP thefts. However,
if an input that leads to a vulnerable state is discovered by
analyzing the obfuscated program it will not go away when the program
is deployed. The attacker in the possession of the critical input will
remain in the position to lead the system to a vulnerable state. The
same is not true for our Drndalo method. In Drndalo, the inputs that
lead to a vulnerable state discovered in the obfuscated version of the
program normally do not lead the deobfuscated version of the program
to a vulnerable state. Nonetheless, the approach of Cousins et al.\ is
proved to be simulation-secure --- the measure of security that we
will describe later in this section --- while Drndalo is not.

Not only that program obfuscation is historically developing to
fulfill multiple purposes but it is also investigated by multiple
research communities. For example, cryptographic research community
gives us the definition of cryptographic obfuscation. We can utilize
this definition to place our Drndalo method in a spectrum of security
and usability of the existing obfuscation solutions.

Barak et al.~\cite{Barak_2012}, in their widely-known work on
feasibility of obfuscation, define obfuscation and the
obfuscator. They intuitively posit an obfuscator $\mathcal{O}$ as an
efficient probabilistic ``compiler'' that takes the plain program $P$
and produces the obfuscated program $\mathcal{O}(P)$. The programs
$\mathcal{O}(P)$ and $P$ are guaranteed to have the same functionality
and to be ``unintelligible'' from each other. In other words, there is
nothing that an adversary can compute from $\mathcal{O}(P)$ and cannot
from only an oracle access to $P$. We refer to such an obfuscator as
simulation-secure. However, the requirements of ``unintelligibility''
and ``same functionality'' allow for certain amount freedom in
interpretation. Thus, Barak et al.\ define the requirements for an
simulation-secure obfuscator by utilizing the concepts of Turing
Machine(TM) and Probabilistic Polynomial-time Turing Machine(PPT) as
are known in complexity-theory, and oracles as known in cryptography.

A TM $\mathcal{O}$ is the obfuscator for the family of TMs
$\mathcal{F}$ iff all the following are true:

\begin{itemize}
\item \textit{(functionality preservation)} For all TMs
  $M \in \mathcal{F}$, $\mathcal{O}(M)$ denotes the obfuscated $M$
  such that it computes the same function as $M$ itself.
\item \textit{(polynomial slowdown)} There exists a polynomial $p$
  such that for any $M \in \mathcal{F}$,
  $|\mathcal{O}(M)| \leq p(|M|)$. Additionally, if $M$ halts within
  $t$ steps, $\mathcal{O}(M)$ must halt in $p(t)$.
\item \textit{(virtual black box)} For all PPT $\mathcal{A}$ and all
  TMs $M \in \mathcal{F}$, there exists a PPT $\mathcal{S}$ that has
  the oracle access to $M$, such that for a negligible function $\phi$
  \begin{equation}
    \label{eq:black_box_obf}
    \lvert {Pr \lbrack {\mathcal{A}(\mathcal{O}(M)) = 1} \rbrack -
      Pr \lbrack {\mathcal{S}^{\langle {M} \rangle}(1^{|M|}) = 1}
      \rbrack} \rvert \leq \phi(|M|)
  \end{equation}
\end{itemize}

Since TMs are intuitively harder to obfuscate than circuits, Barak et
al.\ use
$\exists \mathcal{O}_{TM} \Rightarrow \exists
\mathcal{O}_{circuit}$. By showing that a circuit obfuscator does not
exist, they show that a TM obfuscator does not exist either. However,
it is important to note that this well-accepted conclusion
\textit{does not} rule out the possibility of obfuscation for
\textit{all} programs. For example, Barak et al.\ do not rule out the
obfuscators for finite automata or regular expressions. In fact, Lynn
et al.~\cite{Lynn_2004} have shown that the provable obfuscation for
point functions \textit{do exist} under random oracle model and
construct the provable obfuscations of complex access control
functionalities. Later on, Wee~\cite{Wee_2005} demonstrated that the
random oracle is not necessary and replaced it with a probabilistic
hash function based on a one-way permutation. Cousins et
al.\cite{Cousins_2018} have demonstrated the practicality of virtual
black box obfuscation for the programs of the form
$f(x_1, \ldots x_L) = \bigwedge_{i \in I} y_i$, for
$I \subseteq \lbrack {L} \rbrack$. This programs are referred to as
\textit{conjunctions} and can be used as approximations of classifiers
in machine learning~\cite{Xiao_2015}. The approach
from~\cite{Cousins_2018} achieves substantial performance overhead
reduction when compared to other similar methods. However, it still
does not answer the question of encoding arbitrary valid programs to
conjunctions chosen from a distribution with enough entropy. The
distribution entropy criteria affects the security of obfuscation. In
fact, we are not aware of a cryptographic obfuscation approach that is
suitable to arbitrary programs expressible in some high-level language
such as C.

In contrast to cryptographic obfuscation approaches, our Drndalo
method is not constrained to any subset of programs and accepts all
the valid programs that are expressible in a language supported by the
LLVM frontends. Additionally, the functionality preservation criteria
from~\cite{Barak_2012} is interpreted in a narrower
sense. Specifically, the functionality is preserved only with respect
to the trusted core. However, the security of our method is not
derived from the computational hardness of a mathematical problem and
therefore does not satisfy the virtual black box criteria as stated in
Equation~\ref{eq:black_box_obf}. Notwithstanding, the security of our
method is evaluated against the attack model described earlier in
Section~\ref{sec:attack_model}.

To the best of our knowledge, the only obfuscation technique that
guarantees simulation-secure obfuscation for an arbitrary program is
proposed by Nayak et al.~\cite{Nayak_2017}. They propose a thorough
architecture redesign encompassing scratchpad memories, ORAM
(oblivious RAM), instruction scheduling and context switching. Nayak
et al.\ do achieve a provably simulation-secure obfuscation for an
arbitrary program through their hardware redesign but at a price of an
overhead that spans from 8x to 76x. Our Drndalo, although not
simulation-secure, achieves an overhead of only 5\% on PARSEC
benchmarks.


%% file: future_work.tex
\section{Future Work and Discussion}\label{ref:future}

In previous sections we have demonstrated Drndalo obfuscation and
deobfuscation procedures and discussed some of its weak points.
The technique is shown to have a very low performance overhead in
comparison to other similar methods. Our future work will thus seek
improvements in two main directions:
\begin{enumerate*}[label=\arabic*)]
\item improving security guarantees while keeping performance
  and hardware overheads low,
\item extending our attack model to allow the attacker to monitor
    the execution of the obfuscated program.
\end{enumerate*}

To improve security guarantees we will perform cryptanalysis of our
Drndalo obfuscation under the assumption of a cryptographically secure
hash function. Additionally, we will strive to approach the ideal of
simulation-secure obfuscation from Equation~\ref{eq:black_box_obf}
with regards to our relaxed functionality preservation requirement as
described in Section~\ref{sec:rel_work}.

Our future work will also encompass additional hardening of the
obfuscation phase. In the analysis of the classifier's success rates
we concluded that its accuracy comes from exploiting the compiler's
affinity to use some branch instructions more often than others. As an
improvement to the Drndalo technique, we will refine the compiler's
code generation phase so as to utilize branch instructions more
uniformly. In our investigation, we concluded that such a compiler
code generation phase refinement is attainable. The refinement will
not affect the program's functionality. A uniform utilization of
branch instructions will be achieved through a new register usage
pattern.


%% file: concl.tex
\section{Conclusion}\label{sec:concl}
In this work we explored the space of hardware-software co-designs for
control flow obfuscation. We proposed a compile time obfuscation
technique that relies on a cryptographic hash function with a secret
key and inverts each conditional branch with the probability of
50\%. Only a deobfuscator that holds the secret key can completely
restore the control flow of the original program. Then, we evaluated
both in-software and in-hardware deobfuscation techniques and
demonstrated that in-software techniques incur an unacceptable
performance overhead. Finally, we proposed multiple in-hardware
implementations of deobfuscation phase as extensions to BRISC-V
platform. The evaluation on the PARSEC benchmark singles out our
8-cycle and 16-cycle cached-hash implementations as the best
in-hardware deobfuscation technique with an average performance
penalty of around 2\% and 5\%, respectively.  The added area of the
hashed-cache implementation is dominated by the cache, requiring
approximately 10\% and 20\% more hardware resources compared to the
baseline processor. We evaluate the security of our obfuscation method
by training a classifier to predict whether individual branches are
obfuscated or not. We show that classifiers are able to spot some
statistical regularities in the types of instructions used in plain
and obfuscated binaries, and we propose a way of making Drndalo
obfuscation stealthy for these ML classifiers.